\documentclass[aps,pre,twocolumn,groupedaddress,showpacs,floatfix]{revtex4}
\usepackage{dcolumn}
\usepackage{amsmath}
\usepackage{amssymb}
\usepackage{epstopdf}
\usepackage{graphicx}
\usepackage{bm}
\usepackage[T1]{fontenc}
\usepackage{color}
\usepackage{url}
\usepackage[bf]{subfigure}
\usepackage{rotating}

\begin{document}

\title{Spectra of random networks in the weak clustering regime}

\author{Thomas K. DM. Peron$^1$}
\email{thomas.peron@usp.br}
\author{Peng Ji$^{2}$}
\author{J\"urgen Kurths$^{3,4}$}
\author{Francisco A. Rodrigues$^{1, 5, 6}$}
\affiliation{$^1$Institute of Mathematics and Computer Science, University of S\~ao Paulo, S\~ao Carlos, SP 13566-590, Brazil\\
$^2$Institute of Science and Technology for Brain-inspired Intelligence, Fudan University, Shanghai 200433, PR China\\
$^3$Potsdam Institute for Climate Impact Research (PIK), 14473 Potsdam, Germany\\
$^4$Department of Physics, Humboldt University, 12489 Berlin, Germany\\
$^5$Mathematics Institute, University of Warwick, Gibbet Hill Road, Coventry CV4 7AL, UK\\
$^6$Centre for Complexity Science, University of Warwick, Coventry CV4 7AL, UK}

\begin{abstract}
The asymptotic behaviour of dynamical processes in networks can be expressed as a function of spectral properties of the corresponding adjacency and Laplacian matrices. Although many theoretical results are known for the spectra of traditional configuration models, networks generated through these models fail to describe many topological features of real-world networks, in particular non-null values of the clustering coefficient. Here we study effects of cycles of order three (triangles) in network spectra. By using recent advances in random matrix theory, we determine the spectral distribution of the network adjacency matrix as a function of the average number of triangles attached to each node for networks without modular structure and degree-degree correlations. Implications to network dynamics are discussed. Our findings can shed light in the study of how particular kinds of subgraphs influence network dynamics.
\end{abstract}


\maketitle

\section{Introduction}

Many studies have been devoted to the
investigation of how  intrinsic topological properties of networks interfere in the performance of a given dynamical process~\cite{Arenas08:PR,pastor2015epidemic,rodrigues2016kuramoto}. Examples of such properties are the presence of triangles, degree-degree correlations and modular organization. The main motivation of these studies relies on the fact that the majority of dynamical processes addressed are analytically treatable in networks generated through the configuration model. These networks exhibit different properties from real-world maps, mainly in the limit of a large number of nodes. To overcome this discrepancy between random networks and real-world structures, many network models have been proposed in order to create random networks that mimic properties observed in real-world topologies. For instance, recent random models are able to generate networks presenting transitivity~\cite{newman2009random,miller2009percolation,gleeson2009bond} and distributions of pre-defined subgraphs~\cite{karrer2010random,ritchie2017generation}.

Recent advances in these new random network models have naturally motivated the study of well known dynamical process in these structures, such as in percolation and epidemic spreading~\cite{miller2009percolation,newman2009random,gleeson2009bond,zlatic2012networks}, cascade failure~\cite{hackett2011cascades,gleeson2010how} and synchronization~\cite{peron2013synchronization}. Since the standard configuration model generates networks with locally tree-like structures -- i.e. without loops --, one of the most frequent questions addressed is how dynamics is affected by the presence of triangles (cycles composed by three vertices)~\cite{mcgraw2005clustering,hackett2011cascades,house2011epidemic,miller2009percolation,newman2009random,gleeson2009bond,metz2011spectra}. Although many works tackled dynamical processes on clustered networks, there are still crucial questions unsolved. In particular, many works report that the presence of triangles indeed influences dramatically network synchronization by suppressing the collective behavior between oscillators (see, e.g.~\cite{mcgraw2005clustering,ma_emergence_2009}), whereas
other approaches~\cite{peron2013synchronization} actually show the opposite, i.e. that calculations developed for tree-like networks 
describe quite accurately the onset of synchronization on clustered topologies. This accuracy of tree-based analysis has been verified in other dynamical processes, such as bond and $k$-core percolation, epidemic spreading~\cite{melnik2011unreasonable} and in the Ising model~\cite{yoon2011belief,herrero2015ising}. 

To analyze the source of these contradictory conclusions on the same subject, one should look at the network models used. For instance, 
a common method adopted in the early approaches to model dynamical processes on clustered networks is the 
consideration of stochastic rewiring algorithms~\cite{mcgraw2005clustering,kim2004performance}. However, although the degree distribution is kept fixed, other network properties are modified along the variation on the number of triangles~\cite{mcgraw2005clustering}, including the creation of community 
structure and degree-degree correlations~\cite{mcgraw2005clustering,kim2004performance}. 
In some cases, even small changes in the clustering coefficient are able 
to significantly alter the smallest nonzero Laplacian eigenvalue, a quantity strongly related with network synchronizability~\cite{ma_emergence_2009}.

On the one hand, the consideration of stochastic rewiring algorithms suggests that a higher number of triangles can indeed influence network dynamics, but on the other hand, as previously mentioned, many other network properties end up being significantly changed when the networks are generated. Hence, the isolated influence of triangles on network dynamics is still an open problem. In order to carefully investigate the influence of clustering on dynamical processes, here we study the spectrum of clustered networks disentangling it from other network properties. In particular, we analyze the spectrum of the adjacency matrix of networks that exhibit a non-vanishing clustering coefficient, since many asymptotic properties of critical exponents of several dynamical processes in networks depend on such eigenvalues~\cite{Arenas08:PR,pastor2015epidemic,rodrigues2016kuramoto,dorogovtsev2003spectra}. Thus, analyzing the spectrum of clustered networks enables us to get a general picture of how different is the performance of dynamical processes in these structures compared with locally tree-like networks.

Here we focus on the model proposed independently by Newman and Miller~\cite{newman2009random,miller2009percolation}, which allows generating clustered networks with vanishing assortativity and no modular structure (for results on spectra of regular graphs with cycles of arbitrary lengths c.f.~\cite{metz2011spectra,bolle2013spectra,eckstein2005hopping}). More precisely, here we
develop an analytic method to calculate the spectra of clustered random networks by
generalizing the approach in~\cite{nadakuditi2013spectra} so as to account for the presence of triangles in the networks generated by the model in~\cite{newman2009random,miller2009percolation}. Our theoretical results accurately reproduce the 
positive skewed distribution of eigenvalues of the adjacency matrix. Interestingly, we further reveal that
the largest eigenvalue of this matrix deviates from the value predicted for the traditional configuration model with the same connectivity
by a factor inversely proportional to the average number of triangles. However, we show that this 
difference vanishes as the average degree increases, explaining why critical dynamical properties 
of clustered and locally tree-like networks tend to converge in such a limit. 




\section{Network model}
\label{sec:random_clustered_net}

The model proposed in~\cite{newman2009random,miller2009percolation} consists in setting a sequence of single edges $\{s_i\}$ and a sequence of triangles attached to each node $\{t_{ i}\}$, for $i=1,...,N$. By randomly connecting the stubs of each sequence, we obtain a network in which the conventional degree of a node $i$ is given by $k_i = s_i + 2t_{ i}$, since a triangle contributes with two edges to the node degree. With the joint sequence $\{s_i, t_{ i}\}$ it is possible to define the joint probability distribution $p_{st}$, which is the probability that a randomly selected node is attached to $s$ single edges and $t$ triangles. Moreover, the joint probability distribution $p_{st}$ is related to the conventional degree distribution $p_k$ according to
\begin{equation}
p_k = \sum_{s,t=0}^{\infty} p_{st}\delta_{k,s+2t},
\label{eq:pst}
\end{equation}
where $\delta_{i,j}$ is the Kronecker delta.

Having defined the distribution of single edges and triangles, it is possible to calculate the network transitivity $\mathcal{T}$ as a function of the moments of the distribution $p_{st}$ as
\begin{equation}
\mathcal{T} = \frac{3 \times \mbox{(number of triangles in the network)}}{\mbox{(number of connected triples)}}=\frac{3 N_{\triangle}}{N_3},
\label{eq:transitivity}
\end{equation}
where $3N_\triangle = N\sum_{st} t p_{st}$ and $N_3 = N\sum_k \binom{k}{2}p_k$. This model yields networks in the so-called \textit{weak clustering regime}~\cite{serrano2006clustering}, since the local clustering 
coefficient $c$ of a node with degree $k$ is limited to $c(k)\leq (k-1)^{-1}$.


Here, however, we shall address a variant of the model by Newman and Miller, namely we consider it in the context of the \textit{expected degree configuration model}~\cite{nadakuditi2013spectra,chung2003spectra}. In this paradigm, instead of determining the actual number of connections, sequences $\{s_1,...,s_N \}$ and $\{t_1, ..., t_N\}$ set the expected number of single edges and triangles for each node. Similarly as in the standard expected degree configuration model, the expected number of single edges between nodes $i$ and $j$ is a Poisson distributed random variable 
with mean $s_i s_j/ \sum_q s_q$; and the expected number of triangles formed by the triple $\{i,j,k\}$ is a Poisson random variable with mean $2 t_i t_j t_k /( \sum_q t_q)^2$. The expected degree of node $i$ is obtained by summing over all nodes the probabilities of forming single edges and triangles; i.e. $k_i = \sum_j s_i s_j/\sum_q s_q + \sum_{j,k} 2 t_i t_j t_k /( \sum_q t_q)^2 = s_i + 2t_i$. 

Defined in such a way, in the limit of large degrees, the properties of the expected configuration model for single edges and triangles become identical to the ones of the model proposed in~\cite{newman2009random,miller2009percolation}. It is noteworthy mentioning that a similar extension to the context of expected degrees of the model by Newman and Miller was also recently addressed in~\cite{lim2016phase}.

\section{Spectra of clustered random networks}

In this section we estimate the spectral density $\rho(\lambda)$ of the adjacency matrix $\mathbf{A}$ of the model of 
expected single edges and triangles defined previously. In order to accomplish this goal, we generalize the approach 
in~\cite{nadakuditi2013spectra} in order to take into account the presence of triangles. Our methodology will be as follows: first we calculate the spectral density of the modularity matrix $\mathbf{B}$ defined as
\begin{equation}
\mathbf{B} = \mathbf{A} - \langle \mathbf{A} \rangle, 
\label{eq:modularity}
\end{equation}
whose elements are random variables (not completely independent, as we shall see) with zero mean. Matrix $\mathbf{A}$ is the adjacency matrix and $\langle \mathbf{A} \rangle $ is its ensemble average which, for the network model 
of the previous section, has elements defined as $\langle A_{ij} \rangle = s_i s_j/N \langle s \rangle + 2t_i t_j/N \langle t \rangle$.
After uncovering the eigenvalues of $\mathbf{B}$, we obtain the ones of $\mathbf{A}$ by calculating the contribution of adding  
$\langle \mathbf{A} \rangle $ to the spectrum of $\mathbf{B}$.

The average spectrum over an ensemble of realizations of a random symmetric matrix $\mathbf{{B}}$ with zero-mean off-diagonal elements can be calculated by using the Stieltjes transform of its average resolvent, i.e.~\cite{nadakuditi2013spectra,bai2009spectral,tao2012topics}
\begin{equation}
\rho(\lambda) =  - \frac{1}{N\pi} \textrm{Im} \textrm{Tr} \langle \left(\lambda \mathbf{I} - \mathbf{B}  \right)^{-1} \rangle,
\label{eq:density}
\end{equation}
with the eigenvalue $\lambda$ approaching the real line from above. In order to calculate
the average resolvent in Eq.~(\ref{eq:density}), here we extend the approach in~\cite{nadakuditi2013spectra} 
to clustered random graphs constructed according to the expected degree configuration model described in the previous
section. We start by decomposing a generic $N \times N$ matrix $\mathbf{X}$ in its top 
left $(N-1) \times (N-1)$ matrix (denoted as $\mathbf{X}_n$), bottom right element $x_{nn}$, 
and its last column and row, that is~\cite{nadakuditi2013spectra}
\begin{equation}
\mathbf{X} = \begin{pmatrix}
  \boxed{\begin{matrix} \\ \\ \qquad \mathbf{X}_n \qquad \quad \\ \\ \\\end{matrix}}
  &
  \boxed{\begin{matrix} \\ \phantom{m} \\ \mathbf{a}_n \\ \\ \null \end{matrix}}
  \\
  \boxed{\qquad\mathbf{a}^T_n\qquad} & x_{nn}\rule{0pt}{16pt} \\
          \end{pmatrix}.
\end{equation}    
Consider also the vector $\mathbf{v}_n = \mathbf{X}^{-1}\mathbf{u}$, where $\mathbf{u}=(0,...,0,1)^T$.
By separating $\mathbf{v}_n$ in its first $N-1$ elements and its last element as 
$\mathbf{v}_{n} = (\mathbf{v}_n^{(1)}|v_n)$, one is able to get
\begin{equation}
\begin{aligned}
\mathbf{X}_n \mathbf{v}_n^{(1)} + v_n \mathbf{a}_n = \mathbf{0},\\
\mathbf{a}_n^T \mathbf{v}_n^{(1)} + x_{nn} v_n = 1.
\end{aligned}
\end{equation}
Together, the previous equations give us
\begin{equation}
v_n = [\mathbf{X}^{-1}]_{nn} = \frac{1}{x_{nn} - \mathbf{a}_n^T \mathbf{X}_n^{-1}\mathbf{a}_n}, 
\label{eq:X_-1}
\end{equation}
and 
\begin{equation}
\mathbf{v}_n^{(1)}= - [\mathbf{X}^{-1}]_{nn} \mathbf{X}_n^{-1} \mathbf{a}_n.
\label{eq:v1}
\end{equation}
Following~\cite{nadakuditi2013spectra}, we make the assumption that, in the limit of 
large $N$, the elements $v_n$ are narrowly peaked around their
average value, a condition that is met when the node degrees
are large enough. This allows us to write $\langle [\mathbf{X}^{-1}]_{nn} \rangle$ as
\begin{equation}
\langle [\mathbf{X}^{-1}]_{nn} \rangle = \frac{1}{\langle x_{nn} \rangle - \langle \mathbf{a}_n^T \mathbf{X}_n^{-1}\mathbf{a}_n \rangle}.
\label{eq:avg_X_-1}
\end{equation}
The above expression can also be straigthforwadly generalized to uncover the eigenspectra of Laplacian 
matrices by simply averaging its right-hand side over the distribution of the elements $x_{nn}$~\cite{peixoto2013eigenvalue}.

The strategy now is to set $\mathbf{X} = ( \lambda \mathbf{I} - \mathbf{B} )$ and
calculate the average resolvent through Eq.~(\ref{eq:avg_X_-1}). But before that 
we should evaluate the right term in the denominator of Eq.~(\ref{eq:avg_X_-1}).
For the case of the standard configuration model treated in~\cite{nadakuditi2013spectra}, the calculation of $\langle \mathbf{a}_n^T \mathbf{X}_n^{-1}\mathbf{a}_n \rangle$ is reduced to the sum on the
diagonal terms of $\mathbf{X}^{-1}$ due to the independence of connections, i.e. 
$\langle \mathbf{a}_n^T  \mathbf{X}_n^{-1}\mathbf{a}_n \rangle = \sum_j \langle [\mathbf{X}_n^{-1}]_{jj} \rangle \langle (\mathbf{a}_n)^2_j \rangle$. 
However, for the model of single edges and triangles, the elements of the matrix $\mathbf{X}$ are not completely independent. More precisely, if node $i$ is connected to $j$ and
$j$ is connected to $k$, then there exists a non-null probability that $i$ also shares a link with $k$, since
the network model contemplates the existence of triangles in the limit of large $N$. Thus, 
there must exist some correlation among the elements $A_{ij}$, $A_{jk}$ and $A_{ki}$ of the adjacency matrix. Having said that, we open the sum in Eq.~(\ref{eq:avg_X_-1}) as
\begin{eqnarray}
\langle\mathbf{a}_{n}^{T}\mathbf{X}_{n}^{-1}\mathbf{a}_{n}\rangle & = & \sum_{j}\langle[\mathbf{X}_{n}^{-1}]_{jj}\rangle\langle(\mathbf{a}_{n})_{j}^{2}\rangle\nonumber \\
 &  & +\sum_{j\neq k}\langle(\mathbf{a}_{n}^{T})_{j}[\mathbf{X}_{n}^{-1}]_{jk}(\mathbf{a}_{n})_{k}.\rangle
\label{eq:avg_aXa}
\end{eqnarray}
However, from Eq.~(\ref{eq:v1}) we have that
\begin{equation}
(\mathbf{v}_n^{(1)})_{i}=[\mathbf{X}^{-1}]_{in}=-[\mathbf{X}^{-1}]_{nn}\sum_{j}[\mathbf{X}_{n}^{-1}]_{ij}(\mathbf{a}_{n})_{j}.
\label{eq:v1_open}
\end{equation}
Inserting the previous result into Eq.~(\ref{eq:avg_aXa}) we get
\begin{equation}
\begin{aligned}
&&\langle\mathbf{a}_{n}^{T}\mathbf{X}_{n}^{-1}\mathbf{a}_{n}\rangle = \sum_{j}\langle[\mathbf{X}_{n}^{-1}]_{jj}\rangle\langle(\mathbf{a}_{n})_{j}^{2}\rangle \phantom{aaaaaaaaaaaaaaaaa}\\
 &&-\sum_{j\neq k}\sum_{m}\langle[\mathbf{X}^{-1}]_{kk}[\mathbf{X}_{n}^{-1}]_{jm}\rangle\langle(\mathbf{a}_{n}^{T})_{j}(\mathbf{a}_{k})_{m}(\mathbf{a}_{n})_{k}\rangle, 
\end{aligned}
\label{eq:avg_aXa_2}
\end{equation}
where we have used the independence between $\mathbf{X}^{-1}$ and $\mathbf{a}$~\cite{tao2012topics}. Recalling 
that $(\mathbf{a}_j)_i = - B_{ij}$ we then have that $\langle (\mathbf{a}_j)_i \rangle = - \langle B_{ij}\rangle = - \langle A_{ij}\rangle $. As mentioned previously, the elements $\langle A_{ij}\rangle $ are given by the probability that nodes $i$ and $j$ are conneced and, since the elements $ A_{ij} $ are Poisson
distributed, the variance $\langle A^2_{ij} \rangle $ coincides with the mean $\langle A_{ij} \rangle$.
In the configuration model of the previous section, the nodes can be connected by a single edge or via 
participating in a triangle. It thus follows that $\langle A^2_{ij} \rangle  = s_i s_j/N \langle s \rangle + 2t_i t_j/N \langle t \rangle$. 

To evaluate the second sum in Eq.~(\ref{eq:avg_aXa_2}), we should 
note that $\langle(\mathbf{a}_{n}^{T})_{j}(\mathbf{a}_{k})_{m}(\mathbf{a}_{n})_{k}\rangle = -
\langle A_{nj} A_{mk} A_{kn} \rangle$. The term $\langle A_{nj} A_{mk} A_{kn} \rangle$ is 
equivalent to the probability that there exists simultaneously an edge between nodes $j$ and $n$, $m$ and $k$, and $k$ and $n$. It turns out that the occurrence of connections are independent events, except
concerning the formation of triangles. Therefore,  the only terms that contribute 
to the second sum in Eq.~(\ref{eq:avg_aXa_2}) are $\langle A_{nj} A_{jk} A_{kn} \rangle$; i.e. when $m=j$.
Moreover, $\langle A_{nj} A_{jk} A_{kn} \rangle$ actually consists of the probability that
nodes $n$, $j$ and $k$ share a triangle and its value is given by
\begin{equation}
\langle A_{nj} A_{jk} A_{kn} \rangle = 2 \frac{t_n t_j t_k }{(N \langle t \rangle)^2}.
\label{eq:prob_triangle_njk}
\end{equation}
Substituting the expression for $\langle A^2_{ij} \rangle$ and Eq.~(\ref{eq:prob_triangle_njk}) into Eq.~(\ref{eq:avg_aXa_2})
yields
\begin{equation}
\begin{aligned}
&&\langle\mathbf{a}_{n}^{T}(\lambda\mathbf{I}-\mathbf{B}_{n})^{-1}\mathbf{a}_{n}\rangle  = \frac{s_{n}}{N\langle s\rangle}\sum_{j}\langle[(\lambda\mathbf{I}-\mathbf{B}_n)^{-1}]_{jj}\rangle s_{j}\\
&&  +  \frac{2t_{n}}{N\langle t\rangle}\sum_{j}\langle[(\lambda\mathbf{I}-\mathbf{B}_{n}^{-1}]_{jj}\rangle t_{j}\\
 && + \frac{2t_{n}}{(N\langle t\rangle)^{2}}\sum_{j}\langle[(\lambda\mathbf{I}-\mathbf{B})^{-1}]_{kk}[(\lambda\mathbf{I}-\mathbf{B}_{n})^{-1}]_{jj}\rangle t_{j}t_{k}.
\end{aligned}
\label{eq:avg_aXa_3}
\end{equation}
In the limit $N \rightarrow \infty$, the exclusion of the last row and column of matrix $(\mathbf{\lambda I - B})$ becomes 
negligible and we can thus omit its subscript of $\mathbf{B}$ in Eq.~(\ref{eq:avg_aXa_3}), without loss of generality.  After this consideration, Eq.~(\ref{eq:avg_X_-1}) then reads
\begin{widetext}
\begin{equation}
\langle[(\lambda\mathbf{I}-\mathbf{B}){}^{-1}]_{nn}\rangle=\frac{1}{\lambda-\frac{s_{n}}{N\langle s\rangle}\sum_{j}\langle[(\lambda\mathbf{I}-\mathbf{B}){}^{-1}]_{jj}\rangle s_{j}-\frac{2t_{n}}{N\langle t\rangle}\sum_{j}\langle[(\lambda\mathbf{I}-\mathbf{B})^{-1}]_{jj}\rangle t_{j}-\frac{2t_{n}}{(N\langle t\rangle)^{2}}\left[\sum_{j}\langle[(\lambda\mathbf{I}-\mathbf{B})^{-1}]_{jj}\rangle t_{j}\right]^{2}}.
\label{eq:avg_aXa_4}
\end{equation}
\end{widetext}

In order to simplify the notation, we define the function $\gamma_\lambda(s_n,t_n)$:
\begin{equation}
\gamma_\lambda(s_n,t_n) = \langle [(\lambda \mathbf{I} - \mathbf{B})^{-1}]_{nn} \rangle.  
\label{eq:def_gamma}
\end{equation}
In terms of $\gamma_\lambda(s_n,t_n)$ Eq.~(\ref{eq:avg_aXa_4}) becomes
\begin{equation}
\gamma_\lambda(s_n,t_n) = \frac{1}{\lambda - s_n h_s(\lambda) - 2t_n h_t(\lambda) - 2 t_n h_t^2(\lambda) },
\label{eq:gamma}
\end{equation}
where 
\begin{equation}
\begin{aligned}
h_{s}(\lambda)&=&\frac{1}{N\langle s\rangle}\sum_j s_j\gamma_\lambda(s_j,t_j)\\
h_{t}(\lambda)&=&\frac{1}{N\langle t\rangle} \sum_j t_j \gamma_\lambda(s_j,t_j).
\end{aligned}
\label{eq:h_s_h_t}
\end{equation}
The spectral density $\rho(\lambda)$ is then given by 
\begin{equation}
\rho(\lambda) = - \frac{1}{\pi} \textrm{Im}  g(\lambda), 
\label{eq:rho_g_lambda}
\end{equation}
where
\begin{equation}
g(\lambda) = \frac{1}{N} \textrm{Tr} \langle (\mathbf{\lambda I - B})^{-1}\rangle =   \frac{1}{N} \sum_j \gamma_\lambda (s_j,t_j).
\label{eq:g}
\end{equation}

Thus, by jointly solving Eqs.~(\ref{eq:gamma}) and~(\ref{eq:rho_g_lambda}) one obtains the spectral 
density $\rho(\lambda)$ of the modularity matrix $\mathbf{B}$. In order to obtain 
the spectral density of $\mathbf{A}$, we follow the same argument as in~\cite{nadakuditi2013spectra,benaych2011eigenvalues}. More specifically, 
first we write the adjacency matrix in terms of the modularity matrix  as $\mathbf{A = B + \langle A \rangle}$. 
Second, in order to $\lambda$ be an eigenvalue of $\mathbf{A}$, the equation $\det(\mathbf{\lambda I - (B + \langle A \rangle)})=0$ should be satisfied. However, we also have that
\begin{equation}
\begin{aligned}
\det(\mathbf{\lambda I - (B + \langle A \rangle)}) = \det(\mathbf{\lambda I-B})\phantom{------}\\
\times \det(\mathbf{I}-(\mathbf{\lambda I-B})^{-1}\mathbf{\langle A\rangle}).
\end{aligned}
\label{eq:determinants}
\end{equation}
From Eq.~(\ref{eq:determinants}) we note that $\lambda$ is an eigenvalue of $\mathbf{A}$ and not from $\mathbf{B}$ only
if 1 is an eigenvalue of the matrix $\mathbf{(\lambda I - B)^{-1} \langle A \rangle }$. Therefore, we finally 
obtain the eigenvalues of $\mathbf{A}$ by solving~\cite{benaych2011eigenvalues,peixoto2013eigenvalue} 
\begin{equation}
\det(\mathbf{I}-(\mathbf{\lambda I-B})^{-1}\mathbf{\langle A\rangle}) =0, 
\label{eq:det_IBA}
\end{equation}
where $\langle \mathbf{A} \rangle = \mathbf{ss}^T/N\langle s \rangle + 2\mathbf{tt}^T/N\langle t \rangle$.

Equations~(\ref{eq:gamma}),~(\ref{eq:rho_g_lambda}) and~(\ref{eq:det_IBA}) set a general framework for the calculation 
of the spectrum of the adjacency matrix of networks which 
have distributions of single edges and triangles. In the next
sections we demonstrate the applicability of the method 
in some particular network configurations and discuss 
in more detail how the eigenvalues of $\mathbf{B}$ and $\mathbf{A}$ are
related.

\section{Examples}

Let us treat 
first Poisson random networks with a joint degree distribution as 
\begin{equation}
p_{st} = e^{-\left\langle s \right\rangle}\frac{\left\langle s \right\rangle^{s}}{s!} e^{-\left\langle t \right\rangle}\frac{\left\langle t \right\rangle^{t}}{t!}.
\label{Eq:pst_poisson}
\end{equation}
In the expected degree configuration model for single edges and triangles, the probability distribution (\ref{Eq:pst_poisson})
is translated into the distribution $P(s,t)$ of the expected number of single edges $s$ and triangles $t$ as $P(s,t) = \delta(s - \langle s \rangle ) \delta(t - \langle t \rangle)$, where $\delta(\cdot)$ is the Dirac delta function. 
In the absence of triangles ($\langle t \rangle = 0$), we recover the results of~\cite{nadakuditi2013spectra}; i.e. Eq.~(\ref{eq:gamma}) is reduced to 
\begin{equation}
\gamma_\lambda(s_n) = \frac{1}{\lambda - s_n h_s(\lambda)}.
\label{eq:gamma_just_single_edge}
\end{equation}
For large $N$, we may write $h_s(\lambda)$ as
\begin{equation}
h_s(\lambda) = \frac{1}{\langle s \rangle} \int \frac{s P(s)}{\lambda - s h_s(\lambda)}ds.
\label{eq:hs_just_single_edge}
\end{equation}
Integrating with $P(s) = \delta(s - \langle s \rangle)$ and solving the resulting equation for 
$h_s(\lambda)$, we obtain
\begin{equation}
h_s(\lambda) = \frac{\lambda \pm \sqrt{\lambda^2 - 4\langle s \rangle }}{2\langle s \rangle}.
\label{eq:hs2_just_single_edge}
\end{equation}
Choosing the solution with negative sign, so as to have a positive density, and substituting it into Eq.~(\ref{eq:rho_g_lambda}) leads to the spectral density 
\begin{equation}
\rho(\lambda) = \frac{\sqrt{4\langle s \rangle - \lambda^2}}{2\pi \langle s \rangle}, 
\label{eq:semicircle_law}
\end{equation}
which is the standard semicircle law and coincides with the result obtained in~\cite{nadakuditi2013spectra}. 

\begin{figure}[!tpb]
 \centerline{\includegraphics[width=1\linewidth]{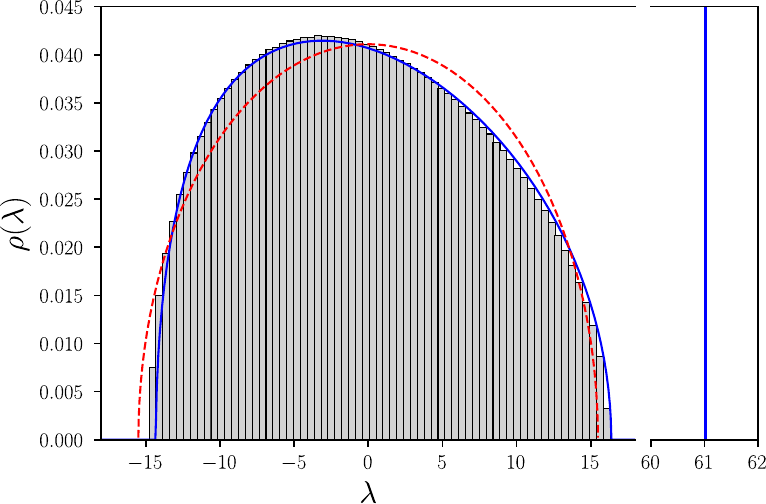}}
  \caption{(Color online)  Spectral density $\rho(\lambda)$ for clustered random networks without singles ($\langle s \rangle = 0$) and distribution of expected number of triangles given by $P(t) = \delta(t - \langle t \rangle)$, with $\langle t \rangle = 30$. Solid blue curve corresponds to the theoretical solution 
  obtained by jointly solving Eqs.~(\ref{eq:ht_cubic_equation}) and~(\ref{eq:g_just_triangles}). Histograms are obtained from numerical calculation considering networks with $N=10^4$ and averages over 100 realizations. Red dashed curve depicts the standard semicircle law for random graphs with the same average degree. Vertical lines on the right mark the position of the leading eigenvalue $\lambda_1$ estimated via Eq.~(\ref{eq:lambda_1_Poisson}) (blue) and numerical calculations (gray, overlapped by the blue line).}
  \label{fig1}
\end{figure}

The maximum clustered topology yielded by the network model of Refs.~\cite{newman2009random,miller2009percolation} is reached
when the degrees are only due to edges that participate in triangles. In order to evaluate
how strongly such motifs modify the network spectrum, it is instructive to address networks 
in which $\langle s \rangle =0$ and $P(t) = \delta(t - \langle t\rangle)$. In this case, Eq.~(\ref{eq:gamma}) becomes
\begin{equation}
\gamma_\lambda(t_n) = \frac{1}{\lambda - 2t_n h_t(\lambda) - 2t_n h_t^2(\lambda) },
\label{eq:gamma_n}
\end{equation}
where $h_t(\lambda)$ is calculated via
\begin{equation}
h_t(\lambda) = \frac{1}{\lambda - 2\langle t\rangle h_t(\lambda)  - 2\langle t \rangle h_t^2(\lambda) }.
\label{eq:ht_just_triangles}
\end{equation}
Rearranging the terms we obtain the cubic equation
\begin{equation}
h_t^3(\lambda) + h_t^2(\lambda) - \frac{\lambda }{2\langle t \rangle}h_t(\lambda) + \frac{1}{2\langle t \rangle} =0. 
\label{eq:ht_cubic_equation}
\end{equation}
Analogously as Eq.~(\ref{eq:ht_just_triangles}), for large $N$, we write Eq.~(\ref{eq:g}) as
\begin{eqnarray}\label{eq:g_t_general}
g(\lambda) & = & \int\frac{P(t)}{\lambda-2th_{t}(\lambda)-2th_{t}^{2}(\lambda)}dt\\
 & = & \frac{1}{\lambda-2\langle t\rangle h_{t}(\lambda)-2\langle t\rangle h_{t}^{2}(\lambda)}.
 \label{eq:g_just_triangles}
\end{eqnarray}
Solving Eq.~(\ref{eq:ht_cubic_equation}) for $h_t$ and inserting the solution in Eq.~(\ref{eq:g_just_triangles}), we get the function $g(\lambda)$, which in turn 
gives us the spectral density via Eq.~(\ref{eq:rho_g_lambda}). Figure~\ref{fig1} shows the yielded result of $\rho(\lambda)$ for clustered
random networks in the absence of single edges ($\langle s \rangle=0 $). As it is seen, the presence of triangles 
leads to a distribution with higher skewness in comparison with the traditional semicircle law (red dashed curve).
This agrees with the known result from spectral graph theory that the skewness coefficient $s_e$ of the spectral distribution $\rho(\lambda)$ should increase with the number of triangles, $s_e = 3N_\triangle /m \sqrt{\left\langle k \right\rangle}$~\cite{mieghem2011graph}, where $m$ is the total number of edges.

\begin{figure}[!tpb]
 \centerline{\includegraphics[width=1\linewidth]{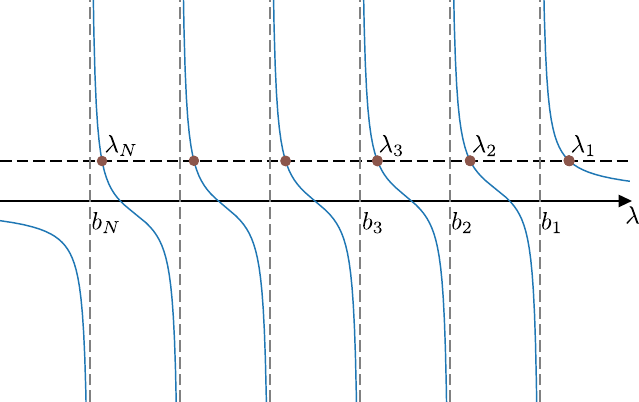}}
  \caption{(Color online)  Determination of the eigenvalues $\lambda_i$ of the adjacency matrix $\mathbf{A}$ in terms of the eigenvalues $b_i$ of the modularity matrix $\mathbf{B}$. Solid blue curves plot the solution 
  of the left-hand side of Eq.~(\ref{eq:solutions_for_lambda_2}). Intersections with 1 (dashed horizontal line) yield the eigenvalues $\lambda_i$. The solutions reveal the interlacing of the eigenvalues of $\mathbf{A}$ and $\mathbf{B}$, i.e. $\lambda_1 \geq b_1 \geq \lambda_2 \geq b_2 \geq ... \geq \lambda_N \geq b_N $.}
   \label{fig2}
\end{figure}

Strictly speaking, the previous equations actually reveal the spectral density of the modularity matrix $\mathbf{B}$; however, as we can see in Fig.~\ref{fig1}, 
this result fits quite accurately the distribution for the adjacency matrix $\mathbf{A}$. The reason for this agreement stems from the 
fact that the eigenvalues of $\mathbf{B}$ and $\mathbf{A}$ are \emph{interlaced}~\cite{nadakuditi2013spectra}. More specifically, consider again networks 
with the highest possible number of triangles ($\langle s \rangle =0$). In this setting, we have that $\mathbf{\langle A \rangle} = 2 \mathbf{tt}^T/N\langle t \rangle$ and Eq.~(\ref{eq:det_IBA}) then gives
\begin{equation}
\frac{2}{N \langle t \rangle} \mathbf{t}^T (\lambda \mathbf{I} - \mathbf{B})^{-1} \mathbf{t} = 1.
\label{eq:solutions_for_lambda_1}
\end{equation}
Relabeling the eigenvalues of $\mathbf{B}$ as $b_i$ and expanding $\mathbf{t}$ in terms of the respective
eigenvectors $\mathbf{x}_i$ of $\mathbf{B}$ yields
\begin{equation}
\frac{2}{N\langle t \rangle} \sum_i \frac{(\mathbf{t}^T\mathbf{x}_i)^2 }{\lambda - b_i } = 1.
\label{eq:solutions_for_lambda_2}
\end{equation}
Figure~\ref{fig2} illustrates the solutions of the equation above. Solid curves represent the left-hand side of Eq.~(\ref{eq:solutions_for_lambda_2}); horizontal,
dashed lines indicate the value 1 of the right-hand side; and the intersection of these curves gives the solutions 
for eigenvalues $\lambda$. Labeling $b_i$ from the largest to the smallest
eigenvalue, and similarly for $\lambda$, from the figure, we have that $\lambda_1 \geq b_1 \geq \lambda_2 \geq b_2 \cdots \geq \lambda_n \geq b_n $. Note also that the eigenvalues of the adjacency
matrix approach the ones of the modularity matrix asymptotically and, therefore, in the limit of large networks the spectra of both matrices coincide. This is valid for the eigenvalues of $\mathbf{A} $ belonging to the bulk, but not for the maximal eigenvalue, which is only bounded from below.

To calculate the largest eigenvalue $\lambda_1$ over an ensemble of
networks, we take the ensemble average of Eq.~(\ref{eq:solutions_for_lambda_1}), i.e. 
\begin{equation}
\frac{2}{N \langle t \rangle} \mathbf{t}^T \langle  (\lambda \mathbf{I} - \mathbf{B})^{-1} \rangle \mathbf{t}= 1, 
\label{eq:int_t_2_gamma_P}
\end{equation}
which, in the limit $N \rightarrow \infty$, can be rewritten as 
\begin{equation}
\frac{2}{\langle t \rangle} \int t^2 \gamma_\lambda(t) P(t) dt = 1.
\label{eq:int_t_3_gamma_P}
\end{equation}
Multiplying Eq.~(\ref{eq:gamma_n}) by $tP(t)$, integrating over $t$ and combining with Eq.~(\ref{eq:int_t_3_gamma_P}) gives
\begin{equation}
(\lambda - 1)h_t(\lambda) - h_t^2(\lambda) = 1, 
\label{eq:equation_for_lambda_1}
\end{equation}
whose solution then yields the largest eigenvalue $\lambda_1$ of $\mathbf{A}$. 

\begin{figure}[!tpb]
 \centerline{\includegraphics[width=1\linewidth]{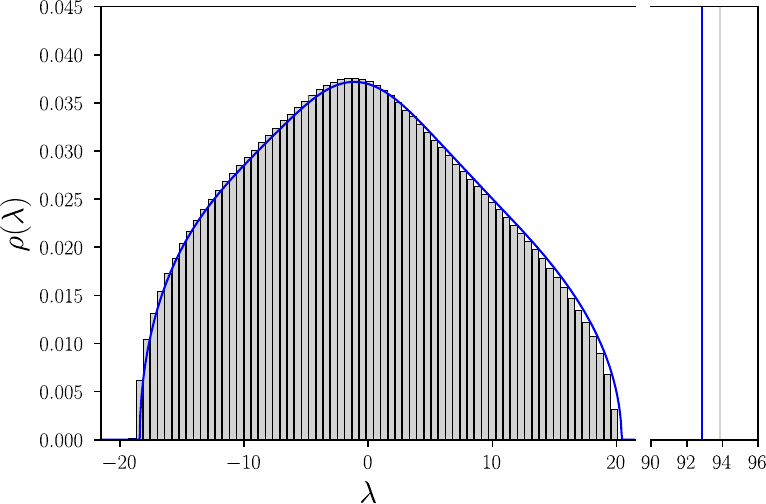}}
  \caption{(Color online)  Spectral density $\rho(\lambda)$ for clustered random networks without singles ($\langle s \rangle = 0$) and distribution of the expected number of triangles given by $P(t) = p_1\delta(t -  t_1 ) + p_2\delta(t -  t_2 )$, with $t_1 = 25$, $t_2 = 50$, $p_1 = 1/4$ and $p_2 = 3/4$. Solid blue curve corresponds to the theoretical solution 
  obtained by solving Eqs.~(\ref{eq:ht_double_t1t2}) and~(\ref{eq:g_t_general}). Histograms are obtained from numerical calculation considering networks with $N=10^4$ and averages over 100 realizations. The blue vertical line on the right mark the position of the leading eigenvalue $\lambda_1$ estimated via Eq.~(\ref{eq:lambda_1_general}) ($\lambda_1 = 92.867...$), while the gray line corresponds to the numerical calculation ($\lambda_1 = 93.867 \pm 0.035$).}
  \label{fig3}
\end{figure}

Consider again the Poisson random graph with the maximal number of triangles ($\langle k \rangle = 2\langle
t \rangle$). By isolating $h_t^2(\lambda)$ in Eq.~(\ref{eq:equation_for_lambda_1}), inserting the result in Eq.~(\ref{eq:ht_just_triangles}) and solving it
for $h_t(\lambda)$, we find that 
\begin{equation}
\lambda_1 =  2\langle t \rangle + 1 + \frac{1}{2\langle t \rangle}.
\label{eq:lambda_1_Poisson}
\end{equation} 
For the 
network ensemble considered in Fig.~\ref{fig1} we obtain with the previous equation $\lambda_1 = 61.0166...$, 
whereas the numerical value is estimated as $\lambda_1 = 61.010 \pm 0.012$, in excellent agreement 
with the value predicted by the theory.  
Interestingly, note that Eq.~(\ref{eq:lambda_1_Poisson}) gives an estimation for $\lambda_1$ very close 
to the well known result for the maximal eigenvalue of Poisson random graphs with the same average degree ($\lambda_1 = \langle k \rangle + 1$)~\cite{nadakuditi2013spectra}.  That is, the largest eigenvalues of clustered and unclustered networks with the 
same average degree only differ by a factor $\langle k \rangle^{-1}$, which vanishes as the network density increases.  This result explains recent findings on the ineffectiveness 
of transitivity in influencing network dynamics in, for instance, epidemic spreading~\cite{house2011epidemic}, 
synchronization~\cite{peron2013synchronization}, and in the Ising model~\cite{herrero2015ising}. 
Critical dynamical properties -- such as epidemic thresholds and critical couplings
for the onset of partial synchronization -- can often be expressed in terms of $\lambda_1$~\cite{pastor2015epidemic,rodrigues2016kuramoto}; 
therefore, one cannot expect to observe differences in the transition points of such dynamical processes in the weak clustering regime 
due to the fact that $\lambda_1$ remains unchanged under the formation of non-overlapping triangles.
Furthermore, it becomes likewise evident in the light 
of Eq.~(\ref{eq:lambda_1_Poisson}) the significant discrepancies that emerge in the dynamics of sparse 
networks subjected to different transitivity levels. This can be seen, for instance, in percolation and cascade failures studied in Refs.~\cite{newman2009random,hackett2011cascades}. It is worth mentioning that the changes in the critical points reported there
cannot be attributed to variations in other network properties, since in the extremal cases 
of the random network model of Refs.~\cite{newman2009random,miller2009percolation} -- only single edges ($\langle t \rangle = 0$) and maximally 
clustered topology ($\langle k \rangle = 2 \langle t \rangle$) -- degree-degree correlations are absent~\cite{huang2013robustness}.



As a last example, we consider a random network without single edges and 
the distribution of the expected number of triangles as
\begin{equation}
P(t) = p_1 \delta(t - t_1) + p_2 \delta(t- t_2).
\label{eq:Pt_double}
\end{equation}
Calculating the function $h_t$ for the previous $P(t)$ gives
\begin{equation}
\begin{aligned}
h_{t}\langle t\rangle[(\lambda-2t_{1}(h_{t}+h_{t}^{2}))(\lambda-2t_{2}(h_{t}+h_{t}^{2}))]=\lambda\langle t\rangle\\-2(h_{t}+h_{t}^{2})t_{1}t_{2},
\end{aligned}
\label{eq:ht_double_t1t2}
\end{equation}
where $\langle t \rangle = t_1 p_1 + t_2 p_2$. Analogously as before, by solving it for $h_t$ and substituting the result in the respective function $g(\lambda)$, we obtain the spectral density $\rho(\lambda)$, which is shown in Fig.~\ref{fig3}. 

In order to estimate $\lambda_1$ for networks following Eq.~(\ref{eq:Pt_double}), let us consider general distributions $P(t)$. By isolating $h_t^2$ in Eq.~(\ref{eq:equation_for_lambda_1}), we express $h_t$ as 
\begin{equation}
h_t(\lambda) = \frac{1}{\langle t \rangle} \int_0^\infty \frac{t P(t)}{\lambda - 2 t \lambda h_t (\lambda) + 2t} dt, 
\label{eq:ht_without_h2}
\end{equation}
which can be expanded as 
\begin{equation}
h_{t}(\lambda)\langle t\rangle\lambda=\int_{0}^{\infty}\sum_{q=0}^{\infty}\left[\frac{2t}{\lambda}(\lambda h_{t}(\lambda)-1)\right]^{q}tP(t)dt.
\label{eq:ht_without_h2_2}
\end{equation}
Considering only the first two terms in the series above we obtain
\begin{equation}
\lambda_1 \approx \frac{2\langle t^2 \rangle }{\langle t \rangle } 
\label{eq:lambda_1_general}
\end{equation}
Figure~\ref{fig3} shows the comparison between the analytical and numerical results for $\lambda_1$ for networks with $P(t)$ according to Eq.~(\ref{eq:Pt_double}). Observe that Eq.~(\ref{eq:lambda_1_general}) is compatible with the corresponding result ($\lambda_1 = \langle k^2 \rangle / \langle k \rangle)$ derived in~\cite{chung2003spectra,nadakuditi2013spectra} for networks generated via the standard expected degree configuration model. Therefore, in the limit of large degrees, we expect clustered networks with general distributions $P(t)$ to exhibit similar critical dynamical properties as unclustered networks with the same average degree, as it is indeed reported in some scenarios~\cite{melnik2011unreasonable,accuracy2012gleeson,peron2013synchronization,yoon2011belief,herrero2015ising}.

\section{Conclusions} 
In summary here we have generalized recent results based on random matrix theory for networks with varying number of triangles, 
enabling us to shed light on the problem about the influence of transitivity on network dynamics. More 
specifically, we have shown how non-overlapping triangles change the network spectrum in the absence of degree-degree 
correlations and modular structure. In particular, for the adjacency matrix $\mathbf{A}$ we have shown that the spectrum distribution 
$\rho(\lambda)$ exhibits positive skewness, in agreement with known results of classical graph theory. Furthermore, 
we showed that the leading eigenvalue of $\mathbf{A}$ of clustered networks exhibits a similar magnitude as in 
unclustered ones, elucidating why critical dynamical properties of the former can often be estimated 
with theories derived for locally tree-like networks~\cite{melnik2011unreasonable,accuracy2012gleeson}.

The ideas presented here together with previous results on spectral graph theory can further motivate developments  in the direction to uncover the 
spectra of networks with more sophisticated subgraph structures. These studies will not only provide a better comprehension of 
network structure, but also contribute to the study of dynamical processes in networks, where the knowledge of the spectrum 
density is required, e.g., in stability analysis and in characterizing the onset of partial synchronization of phase oscillators.
\vskip -0.5cm

\section*{Acknowledgments}
\vskip -0.3cm
TKDMP acknowledges FAPESP (No. 2016/23827-6). PJ is Sponsored by Natural Science Foundation of Shanghai, Shanghai Pujiang Program and by NSFC (11701096). FAR acknowledges the Leverhulme Trust, CNPq (Grant No. 305940/2010-4) and FAPESP (Grants No. 2016/25682-5 and grants 2013/07375-0) for the financial support given to this research. JK would like to acknowledge IRTG 1740 (DFG and FAPESP) for the sponsorship provided. TKDMP further thanks J. A. M\'endez-Berm\'udez and T. Kittel for useful discussions, and Y. Moreno for giving comments on a earlier version of the manuscript.

\bibliographystyle{apsrev}
\bibliography{paper}

\end{document}